\documentclass[12pt,a4paper]{article}
\usepackage{epsfig}         
\usepackage{citemw}
\usepackage{times,mathptm}  

\def\miss{\vspace{\baselineskip}}   

\begin{document}

\vspace*{34mm}
\noindent
{\bfseries\large SMALL-CAPACITANCE JOSEPHSON JUNCTIONS: 
\vspace{0.5\baselineskip}           
\\ ONE-DIMENSIONAL ARRAYS AND SINGLE JUNCTIONS 
\vspace{0.5\baselineskip}
}\\
\par\miss\miss

\noindent
\hspace{24mm}              
\begin{minipage}{5in}
MICHIO WATANABE

\vspace{3mm}
Semiconductors Laboratory, RIKEN (The Institute 
of Physical and Chemical Research), 2-1 Hirosawa, 
Wako-shi, Saitama 351-0198, Japan

\vspace{10mm}
DAVID B. HAVILAND

\vspace{3mm}
Nanostructure Physics, The Royal Institute of 
Technology (KTH),\\ 
SCFAB, Roslagstullsbacken 21, 106 91 Stockholm, Sweden
\end{minipage}
\par\miss\miss


\section{INTRODUCTION}
Small-capacitance superconducting tunnel junctions provide 
an ideal system for studying 
the interplay between the Josephson phase 
and the charge on the junction electrode,   
which are quantum mechanically conjugate to each other.  
This interplay can be probed through the superconductor-insulator (SI)
transition~\cite{Son97}, which is a quantum phase transition occurring 
at $T=0$.  The SI transition has been 
extensively studied in two-dimensional (2D) systems.    
Experiments have been carried out on granular~\cite{Jae89,Heb90,Bar93,Wat97} 
and homogeneous films~\cite{Liu93,Val92}.  
Theoretical studies have modeled the films as 2D arrays of 
small-capacitance Josephson junctions 
(JJs)~\cite{Cha91,Kam93,Sim93,Ott93,Wal94}, 
and experiments with such arrays have also been reported~\cite{Gee89,Che95}.  
In 1D, however, experimental study of the SI transition in JJ arrays 
has been less extensive, while the theoretical investigation has been 
done~\cite{Son97,Cha91,Bra84,Kor89,Bob92,Odi96,Gla97,Choi98}.  
Experimental data on long and narrow films are available~\cite{Sha93,Her96}.  
In contrast to films, JJ arrays can be fabricated with a high degree of 
uniformity, and the parameters of interest in the theory can be measured.  
Furthermore, one can design a JJ array in such a way that one of the 
important parameter in the theory, the Josephson coupling energy $E_J$ 
between adjacent islands can be tuned {\itshape in situ}~\cite{Cho98,Hav00}.  
The other important parameters are 
the charging energy associated with the junction capacitance, 
$E_C\equiv e^2/2C$, and the stray capacitance of each island to the ground, 
$E_{C_0}\equiv e^2/2C_0$.  Depending on the values of these parameters 
either superconducting or insulating behavior is expected 
for an array with infinite length.  

In long arrays, it is possible to observe a well developed Coulomb 
blockade~\cite{Ave91} for Cooper pairs in the current-voltage ($I$-$V$) 
characteristics, even when the Josephson energy is dominant, $E_J \geq E_C$.  
Such a Coulomb blockade is extremely interesting because our usual notions 
about phase coherence in the sense of the Josephson effect do not apply.  
The phase of the superconducting state is washed out by strong quantum fluctuations, 
and the number of Cooper pairs on the island becomes well defined.  Nevertheless, 
the large Josephson coupling causes the potential associated with one excess 
Cooper pair to spread out, and in this sense the single excess Cooper pair 
becomes delocalized.  In the Coulomb-blockade state, the single excess Cooper 
pair can be described as a charge soliton~\cite{Ben89,Hav96PRB}, 
which is dual to the Josephson fluxon of 1D parallel arrays.  
The charge-soliton length, $\lambda_s=2e/2\pi C_0V_c$, gives the length scale 
over which the potential is screened.  Here $V_c$ is the critical 
voltage for Cooper-pair tunneling in a single junction, which is a function 
of the ratio $E_J/E_C$~\cite{Lik85}.  For $E_J \geq E_C$, $V_c$ is 
reduced exponentially to zero as $E_J$ increases.  
This weakening of the Coulomb blockade causes $\lambda_s
\rightarrow\infty$, and we expect the insulating state of the array to eventually 
give way to superconductivity when $E_J \gg E_C$.  In Sec.~\ref{sec:1D}, we describe 
experimental data which display this evolution of the insulating state as $E_J$ 
is tuned {\itshape in situ}.  We interpret the data qualitatively in terms 
of a theoretical model for a $T=0$ quantum phase transition~\cite{Son97}.

We have also used the 1D JJ arrays to bias a single Josephson junction in order 
to control the electromagnetic environment for the single junction~\cite{Wat01PRL}.  
In single junctions, experimental observation of Coulomb blockade has been considered 
to be extremely difficult because a high-impedance environment is necessary, 
and special care should be taken with the measurement leads~\cite{Ave91}.  
For this reason, thin-film resistors~\cite{Hav91} and tunnel-junction 
arrays~\cite{Gee90,Shi97} were employed for the leads, and an increase of 
differential resistance around $V=0$ was reported.  
In contrast to the earlier works~\cite{Hav91,Gee90,Shi97}, our leads 
are tunable, and we can therefore study the {\itshape same} single junction 
in different environments.  
We show that the $I$-$V$ curve of the single junction is indeed sensitive 
to the state of the environment.  Furthermore, we can induce 
a transition to a Coulomb blockade in the single junction 
when the zero-bias resistance of the JJ arrays is much higher 
than the quantum resistance $R_K\equiv h/e^2\approx26$~k$\Omega$.  
In addition to Coulomb blockade, we have clearly observed a region 
of negative differential resistance in the $I$-$V$ curve.  
The negative differential resistance appears as a result of coherent 
tunneling of single Cooper pairs according to the theory of 
current-biased single Josephson junctions~\cite{Ave91,Sch90}.  
Based on the theory, we have calculated the $I$-$V$ curves 
numerically.  The measured $I$-$V$ is consistent with the 
numerical calculation.   

\section{EXPERIMENT}
\subsection{Sample fabrication and characterization}
\begin{figure}
\begin{center}
\epsfig{file=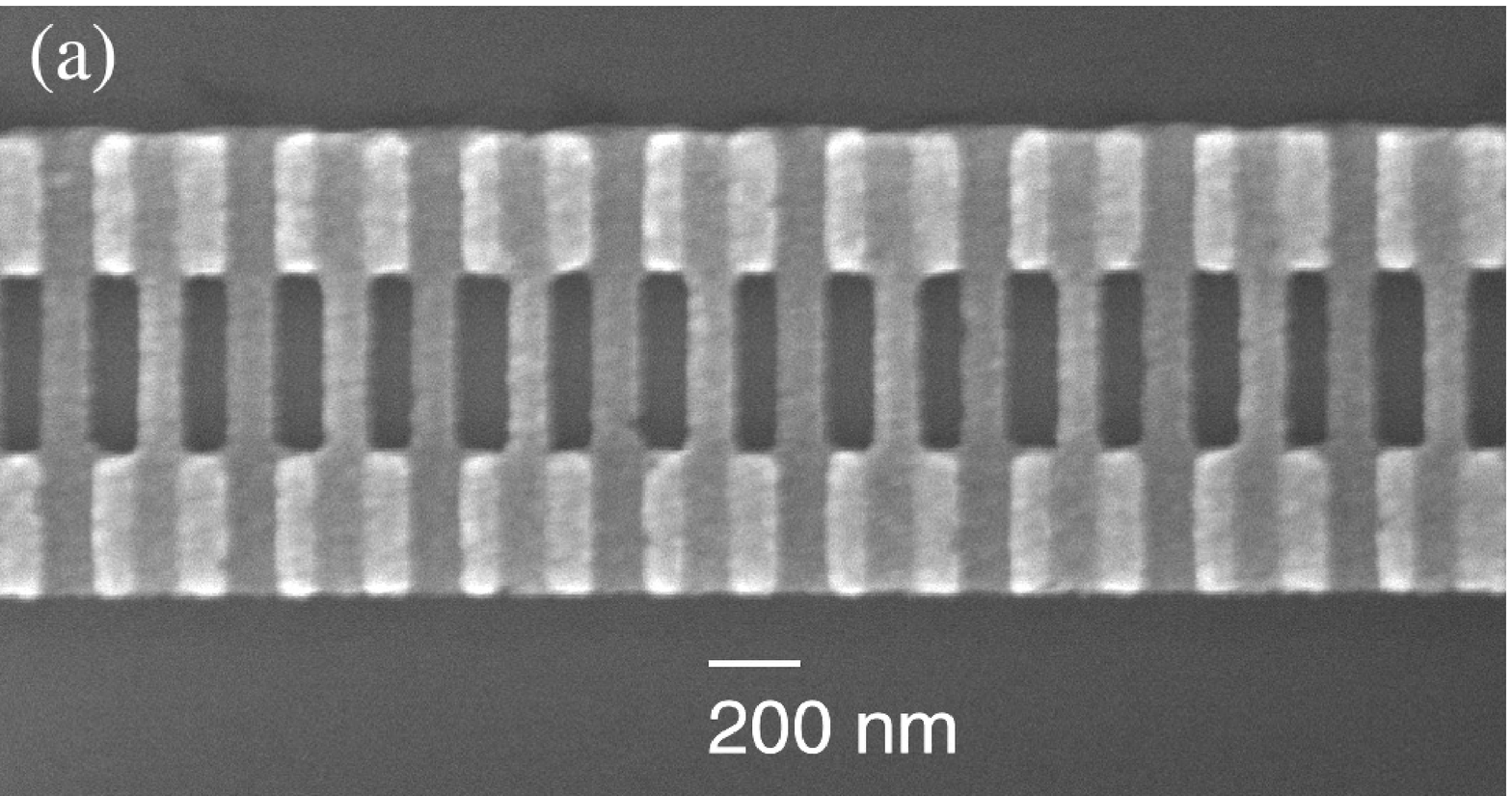,width=.6\columnwidth
}

\miss

\epsfig{file=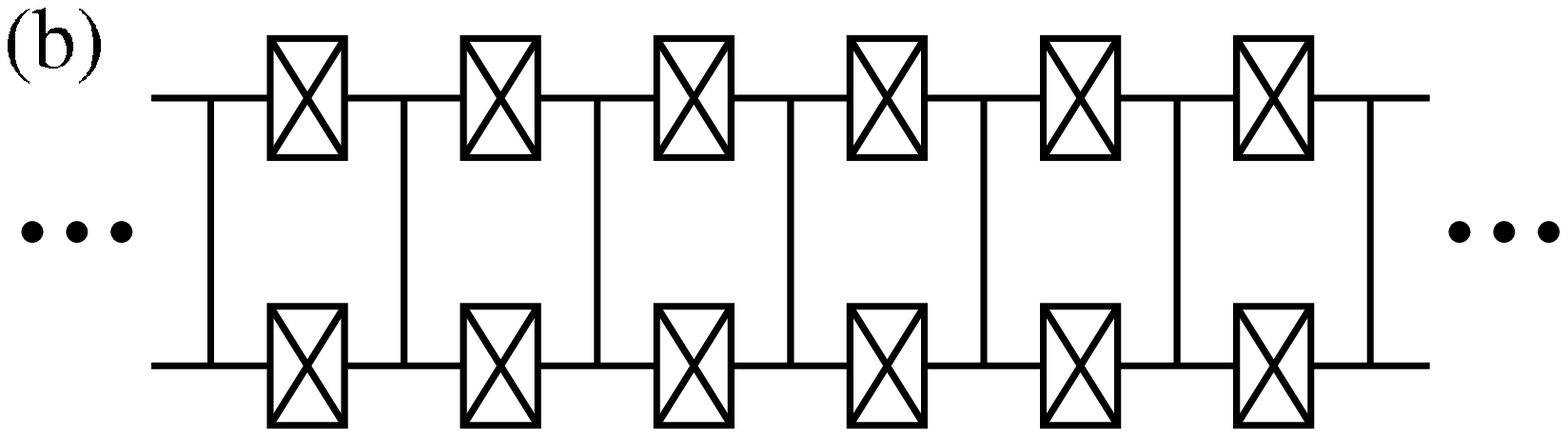,width=.6\columnwidth,
bbllx=53,
bblly=654,
bburx=550,
bbury=793,
clip=,
angle=0
}
\end{center}
\vspace{-.7\baselineskip}
\caption
{
One-dimensional array of small-capacitance dc SQUIDs.  
(a) Scanning electron micrograph.  
(b) Schematic diagram.  
}
\label{fig:array}
\end{figure}

The 1D JJ arrays were fabricated on a SiO$_2$ substrate with electron-beam 
lithography and a double-angle-evaporation technique~\cite{Hav96JVST}.  
The arrays are made of Al with an Al$_2$O$_3$ tunnel barrier, 
and each of the Al electrodes in the array is connected 
to its neighbors by two junctions in parallel, thus forming 
a superconducting quantum interference device (SQUID) 
between nearest neighbors.  
Figure~\ref{fig:array} shows a scanning electron micrograph 
of a section of an array and the schematic diagram.  
The advantage of the SQUID geometry is that we can control 
the effective $E_J$ by applying an external magnetic field, $B$, 
perpendicular to the substrate, 
\begin{equation}
E_J=E_{J0}\left|\cos\left(\pi\,\frac{\,BA_{\rm loop}\,}{\Phi_0}\right)\right|, 
\end{equation}
where $A_{\rm loop}$ is the effective area of the SQUID loop 
and $\Phi_0\equiv h/2e = 2\times10^{-15}$~Wb 
is the superconducting flux quantum.  
Because the geometrical inductance of the SQUID loop, 
$L_0\ll \Phi_0/2\pi I_{c0}$, an external magnetic field 
creates a phase shift so that the critical current 
between nearest neighbors is modulated in a periodic way.  
Here, 
\begin{equation}
E_{J0}\equiv\left(\frac{\Phi_0}{\,2\pi\,}\right)I_{c0}  
\end{equation}
and 
\begin{equation}
I_{c0}\equiv\frac{\pi\Delta_0}{\,2eR_n\,}
\end{equation}
is the Ambegaokar-Baratoff critical current~\cite{Amb63}, which is calculated 
from the superconducting energy gap, $\Delta_0$~($=0.2$~meV for Al), 
and normal-state tunnel resistance of the junction, $R_n$.  
Henceforth, we will refer to the lumped 
SQUID as an effective junction with a tunable $E_J$, and a fixed charging 
energy $E_C\equiv e^2/2C$, where $C$ is the sum (parallel combination) of 
two junction capacitances.  
 
There are a couple of ways to obtain $R_n$.  The resistance of the array divided 
by the number of junctions, $N$, measured above the superconducting 
transition temperature, $T_c$~($=1.2$~K for Al), or in a magnetic field 
strong enough to completely suppress the superconductivity 
($>\,$0.1~T for Al), is $R_n$ by definition.     
It is also possible to find $R_n$ by taking the slope 
of the $I$-$V$ curve at high bias, $V>N(2\Delta_0/e)$.   
The capacitance $C=c_sA$ is estimated  
from the junction area $A$, where the specific capacitance $c_s$ 
is on the order of $10^2$~fF/$\mu$m$^2$ as we will see later 
in Sec.~\ref{sec:SJ}.\ref{subsec:comp}.  
Another important parameter of the array is the capacitance of each electrode 
to the ground, $C_0\sim10$~aF\cite{Hav01}, which depends on $N$ logarithmically.  
In Sec.~\ref{sec:1D}, we will discuss three arrays with nominally 
identical junction parameters [$R_n=4.9$~k$\Omega$,  
$A=(0.4\times0.1$~$\mu$m$^2)\times2$, 
and $A_{\rm loop}=0.7\times0.2$~$\mu$m$^2$], 
but having a different $N$: 255, 127, and 63.  

\begin{figure}
\begin{center}
\epsfig{file=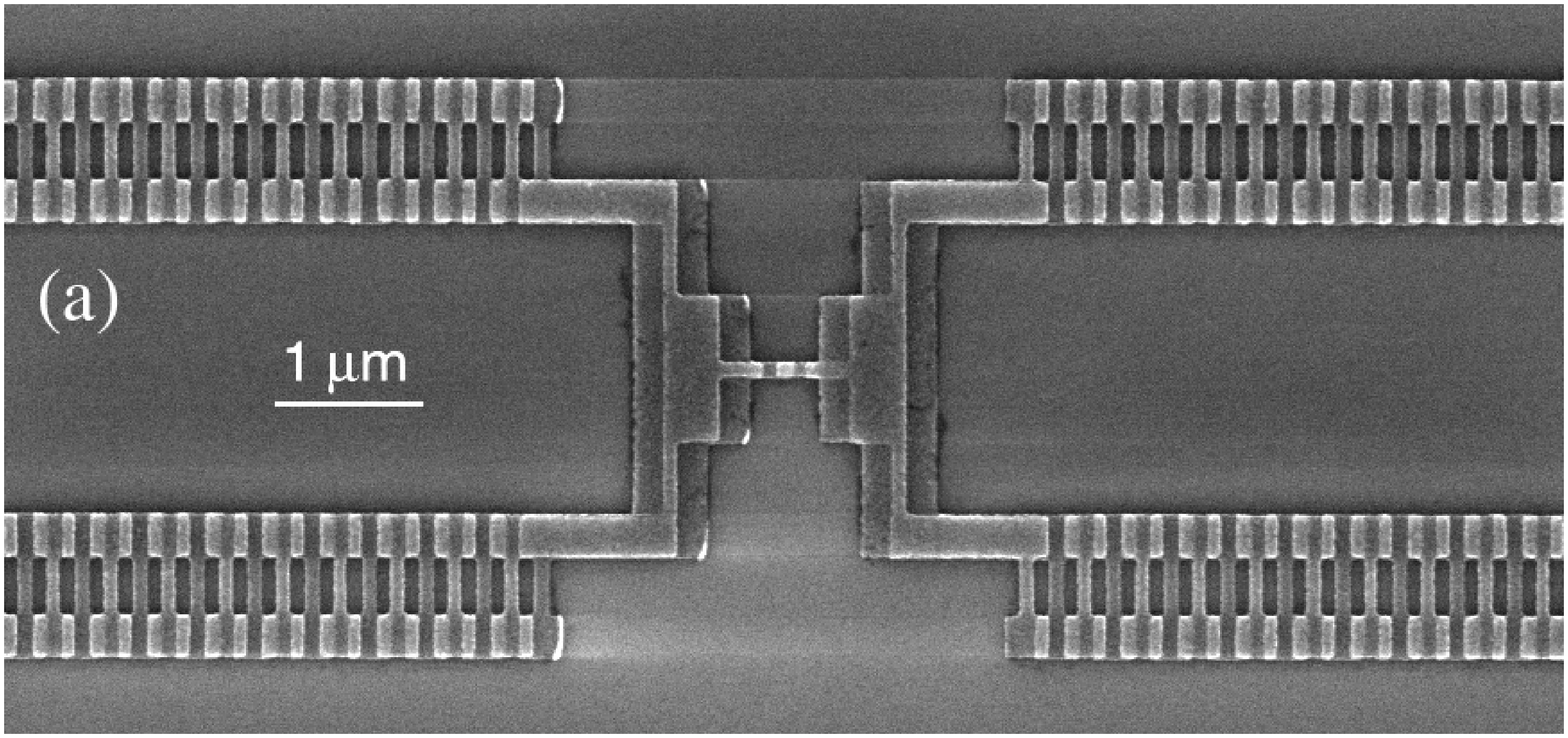,width=.6\columnwidth
}

\miss

\epsfig{file=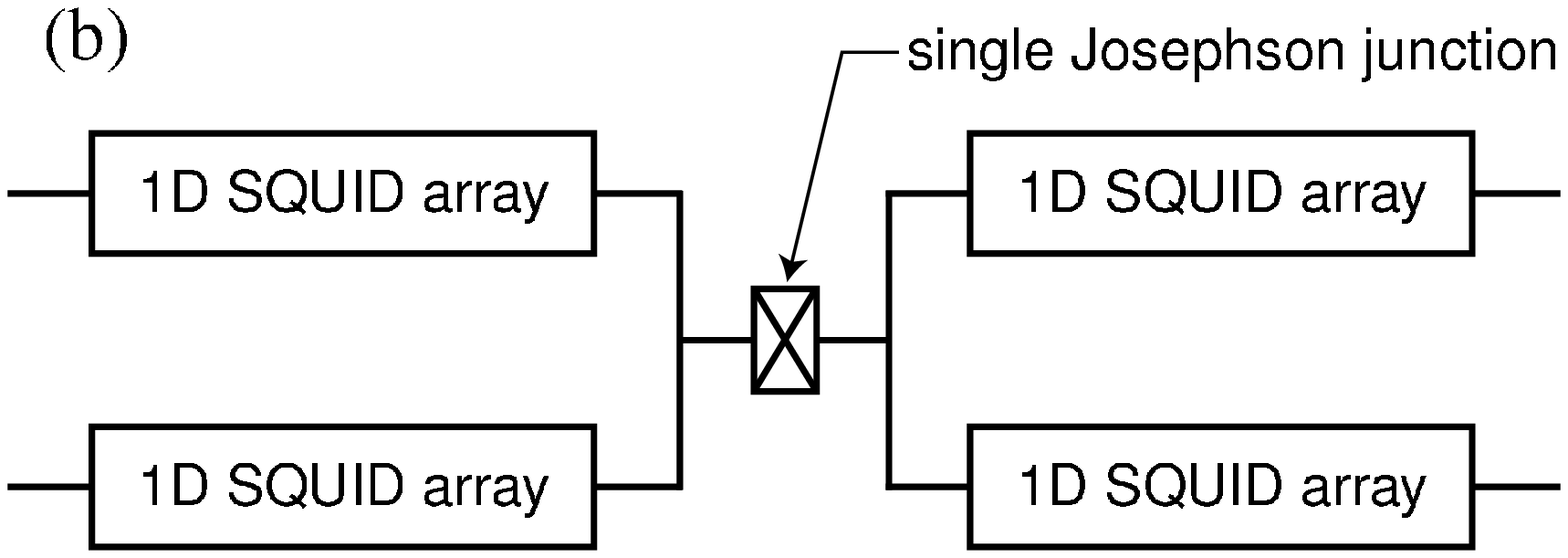,width=.6\columnwidth,
bbllx=41,
bblly=563,
bburx=534,
bbury=737,
clip=,
angle=0
}
\end{center}
\vspace{-.7\baselineskip}
\caption
{
Single Josephson junction biased with arrays 
of small-capacitance dc SQUIDs.  
(a) Scanning electron micrograph.  
(b) Schematic diagram.  
}
\label{fig:sJJ}
\end{figure}

The samples for Sec.~\ref{sec:SJ}.\ref{subsec:SJwS} and \ref{subsec:comp} 
are single Josephson junctions biased with the SQUID arrays.  
(See Fig.~\ref{fig:sJJ}.) 
The single junction, which is in the center of Fig.~\ref{fig:sJJ}a, 
has an area of 0.1$\times$0.1~$\mu$m$^2$.  On each side of 
the single junction there are two leads enabling four-point 
measurements of the single junction.  
A part of each lead close to the single junction consists 
of the SQUID array with $A=(0.3\times0.1$~$\mu$m$^2)\times2$, 
and $A_{\rm loop}=0.7\times0.2$~$\mu$m$^2$.      
The samples are characterized by $R_n$, $N$, and  
the normal-state tunnel resistance of the single junction, $r_n$.   

\subsection{Low-temperature measurements}
The $I$-$V$ curves and the zero-temperature resistance were measured 
in a $^{3}$He-$^{4}$He dilution refrigerator at $0.02-1$~K.  
The temperature was determined by measuring the resistance 
of a ruthenium-oxide thermometer~\cite{Wat01Cryo} fixed at the mixing chamber.
Special care was taken to filter the sample from high-frequency electromagnetic 
radiation~\cite{Hav96JVST}.  
The preamplifier stage of our measurement scheme was specially designed 
for the high resistances associated with the Coulomb blockade.  

The temperature dependence of the zero-bias resistance (Fig.~\ref{fig:Cfig3}) 
was determined with lock-in technique at 13~Hz~\cite{Cho98}.  
The power dissipation was kept at 10$^{-16}$~W, which was just large enough 
to yield a detectable signal, and at the same time, small enough to probe 
the ``linear" response.  

The $I$-$V$ curve of the single junction (Fig.~\ref{fig:21A4SJ}) were 
measured in a four-point configuration, where the potential difference 
was measured through one pair of SQUID-array leads with a high-input-impedance 
instrumentation amplifier, and through the other pair of SQUID-array leads, 
the bias was applied and the current was measured with a current 
preamplifier~\cite{Wat01PRL}.  
When the voltage drop at the SQUID arrays 
was much larger than that at the single junction, 
the single junction was practically current biased.  
The SQUID arrays could be measured in a two-point configuration 
(same current and voltage leads) on the same side of the single 
junction.  Note that the two arrays are connected in series 
and that current does not flow through the single junction.  

\section{SUPERCONDUCTOR-INSULATOR TRANSITION\\ IN ONE-DIMENSIONAL ARRAYS}
\label{sec:1D}
\subsection{Current-voltage characteristics and the zero-bias resistance}
\begin{figure}
\begin{center}
\epsfig{file=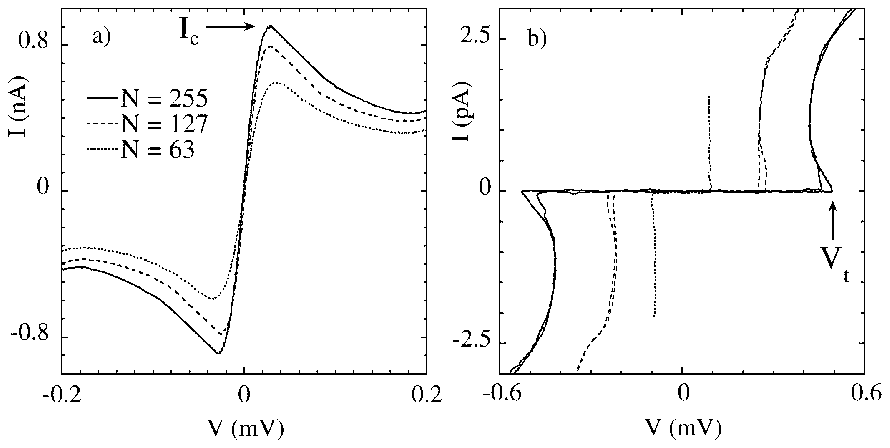,width=\columnwidth
}

\miss\miss

\epsfig{file=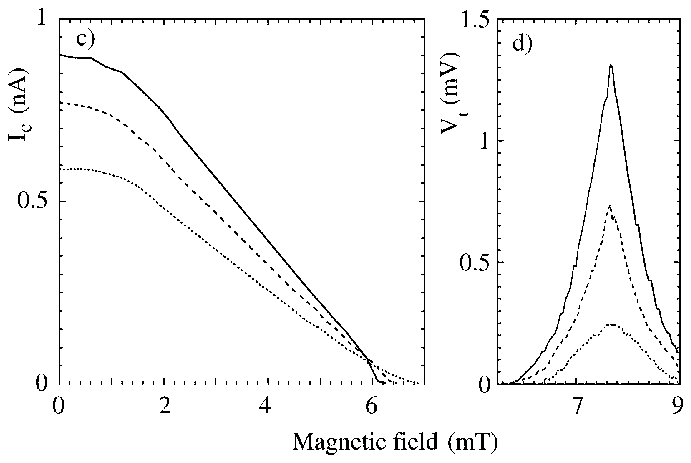,width=.8046\columnwidth
}
\end{center}
\caption
{
Dependence of the $I$-$V$ curves on the array length $N$, at $T=0.05$~K.  
(a) The $I$-$V$ curves at $B=0$ showing Josephson-like behavior and 
the critical current $I_c$.  (b) The $I$-$V$ curves at $B=7.1$~mT 
showing the Coulomb blockade of Cooper-pair tunneling and the threshold 
voltage $V_t$.  (c) The magnetic field dependence of $I_c$.  
(d) The magnetic field dependence of $V_t$.  
}
\label{fig:Chowfig2}
\end{figure}

Figure~\ref{fig:Chowfig2}a shows the $I$-$V$ curve of the three arrays 
at zero magnetic field.  
The arrays were made on the same chip, in the same vacuum cycle, 
using masks written to the same dimensions.  
Thus, all junctions in each array should be identical.  
The arrays are not truly superconducting, and there is actually a slope on the 
``zero-voltage" branch of the $I$-$V$ curve, which gives a finite resistance.  
Furthermore, the observed ``critical currents," i.e., the first local current maximum  
at $\approx0.03$~mV, are only 1\% of the classical Ambegaokar-Baratoff value.  
This critical current shows a clear dependence on the array length.  
The longer the array, 
the lager the critical current, indicating that superconducting behavior is favored 
in the longer array.  As $E_J$ is suppressed below $E_{J0}$ with an externally applied 
magnetic field, the measured critical current of each array is reduced, and the 
resistance on the ``zero-voltage" branch increases.  
Figure~\ref{fig:Chowfig2}c shows 
the magnetic-field dependence of the critical current.  In the neighborhood of $B_c 
=5.8$~mT, the curves in Fig.~\ref{fig:Chowfig2}c cross one another, so that for 
$B>B_c$, the longer the array, the smaller the critical current.   

Figure~\ref{fig:Chowfig2}b shows the $I$-$V$ curve of the three arrays 
at $B=7.1$~mT ($>$$B_c$).  Here we see a new type of behavior which is dual 
to the $B<B_c$ behavior.  The $I$-$V$ curve is characterized by a zero-current 
state for voltage below a threshold voltage, where the array switches 
to a finite current state.  
The magnetic-field dependence of the threshold voltage for $B>B_c$ 
is shown in Fig.~\ref{fig:Chowfig2}d for the three arrays.  
We see that the longer the array, the larger the threshold voltage, 
indicating that insulating behavior is favored in the longer array.  

\begin{figure}
\begin{center}
\epsfig{file=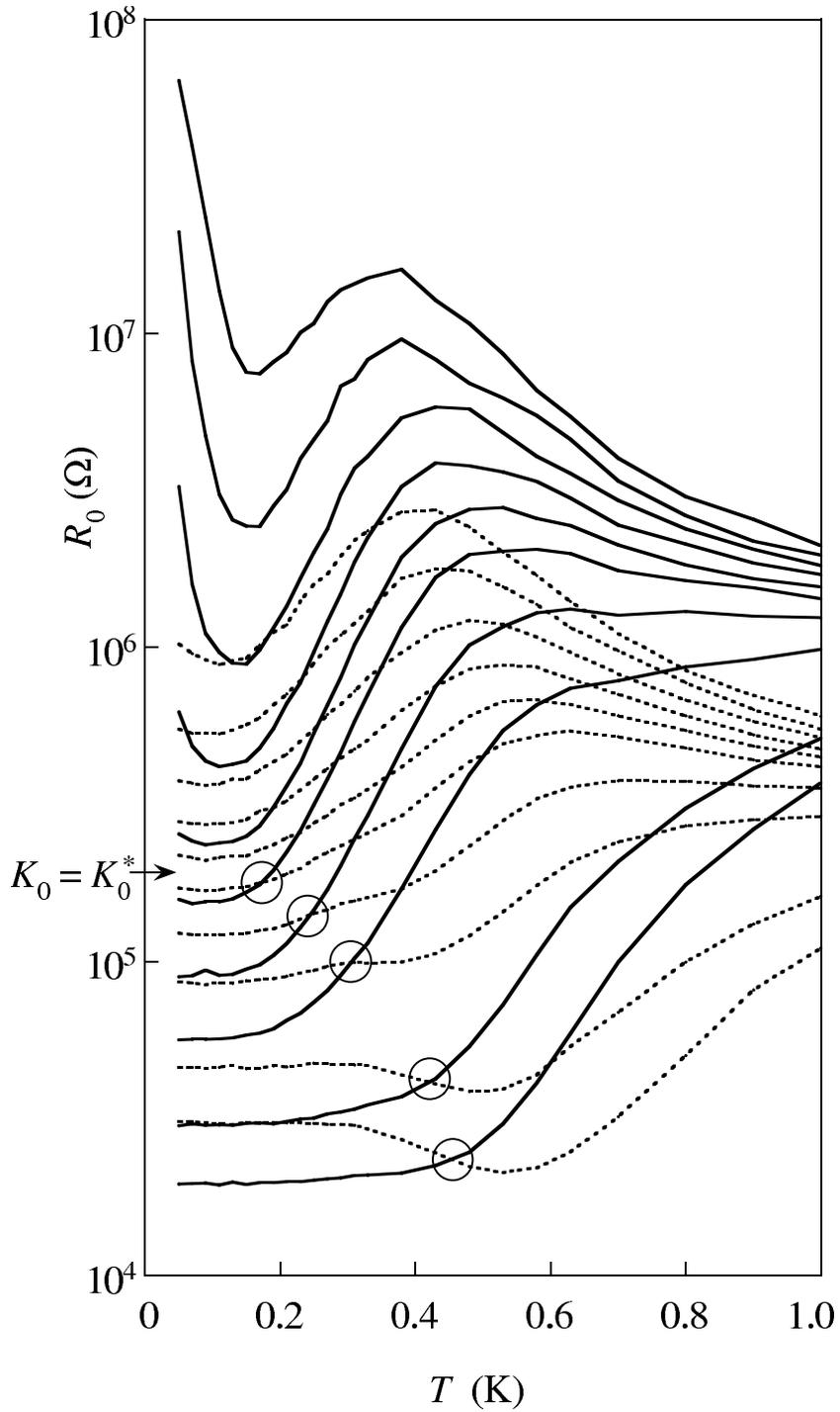,width=0.7\columnwidth,
bbllx=79,
bblly=63,
bburx=500,
bbury=778,
clip=,
angle=0
}
\end{center}
\caption{Zero-bias resistance vs. temperature for two arrays having different 
number of junctions, $N$, but otherwise identical parameters.  The set of 
solid curves are for $N=255$ and the dashed curves are for $N=63$, taken at 
the same magnetic fields between 0 and 7~mT.  The open circles show where the 
measurements on the two arrays at the same magnetic field cross.  At the 
magnetic field where $K_0=K_0^*$, the two arrays have the same 
$T\rightarrow0$ resistance, which is presumably independent of $N$.  
}
\label{fig:Cfig3}
\end{figure}

Figure~\ref{fig:Cfig3} shows the temperature dependence of the 
zero-bias resistance, $R_0(T)$, taken at the same magnetic fields 
(same $E_J$) for two arrays of different length: $N=255$ (solid) 
and 63 (dashed). 
Each set of the curves shows qualitatively similar behavior.  
At zero magnetic field, as the temperature is lowered $R_0$ decreases to a value 
which is temperature independent.  As the magnetic field is increased, 
the resistance of this ``flat tail" increases, until it reaches a critical value, 
where $R_0(T)$ curves make a sharp turn to increasing resistance as $T\rightarrow0$.  
Further increasing of the magnetic field drives the array into the 
insulating state, where $R_0$ increases rapidly as $T\rightarrow0$.  

If we examine the bottom two curves in each set of Fig.~\ref{fig:Cfig3}, 
we can see that at high temperatures, the 63-junction array has a smaller resistance 
than the 255-junction array, as expected for a classical resistor.  However, 
at low temperatures, the resistance of the 63-junction array becomes {\itshape larger} 
than the 255-junction array.  This increasing of the resistance for shorter arrays 
is a clear sign that quantum fluctuations are responsible 
for the measured resistance~\cite{Son97}.  
The open circles in Fig.~\ref{fig:Cfig3} indicate the crossing points where $R_0$ 
is the same for two different lengths at the same magnetic field.  This crossing point 
moves towards $T=0$ as the magnetic field is tuned to the critical point 
$K_0^*$.  At the critical point, the $T\rightarrow0$ resistance due to quantum 
fluctuations, is independent of the array length.   

We have seen a magnetic-field-tuned transition from Josephson-like behavior 
to Coulomb blockade, which may be called a superconductor-insulator transition, 
through the $I$-$V$ curve (Fig.~\ref{fig:Chowfig2}) and the temperature 
dependence of $R_0$ (Fig.~\ref{fig:Cfig3}).  Moreover, the sharpness 
of the transition is strongly influenced by the length of the array.    
We can find qualitative explanation for this length dependence 
in a theoretical model of a quantum phase transition, which will be 
discussed in the following subsection.    

\subsection{Mapping to the {\itshape XY} model}
\begin{figure}
\begin{center}
\epsfig{file=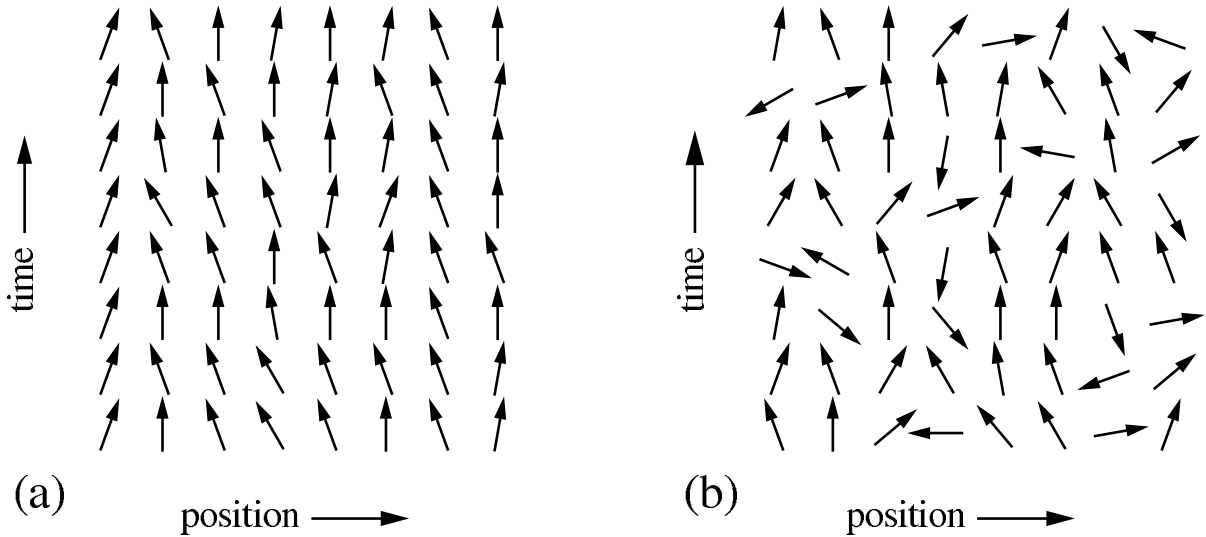,width=\columnwidth,
bbllx=72,
bblly=564,
bburx=530,
bbury=767,
clip=,
angle=0
}
\end{center}
\vspace{-.7\baselineskip}
\caption
{
Typical path or time history of a one-dimensional Josephson-junction 
array in (a) the superconducting phase and in (b) the insulating phase, 
respectively.  The orientation of the arrows indicates the phase 
angles of the superconducting order parameter on the metallic elements 
connected by the Josephson junctions.  
}
\label{fig:XYmodel}
\end{figure}

The SI transition can be described in an elegant theoretical 
framework as a quantum phase transition \cite{Son97}.  In these models, one can 
describe how a $T=0$ property of a macroscopic quantum system with many degrees 
of freedom, will change as the complementary energies in the Hamiltonian 
of the system are adjusted.  One can calculate the linear response, 
which in our case is the zero bias-resistance $R_0$, resulting from quantum 
fluctuations of the degrees of freedom.  Within this framework, 
our 1D quantum system of Josephson junctions is mapped to the classical $XY$ 
model of (1+1)D as sketched in Figs.~\ref{fig:XYmodel}a and \ref{fig:XYmodel}b, 
the extra dimension being imaginary time, $i\hbar/k_BT$.  
Note that the role of temperature for the quantum system is to set the ``size" 
of the system in the imaginary-time dimension.  
The (1+1)D classical $XY$ model exhibits a Berzinski-Kosterlitz-Thouless phase 
transition~\cite{Ber71,Kos73}, from a disordered state (free vortices) 
to an ordered state (bound vortex pairs) as the strength of the dimensionless 
coupling constant, $K_0$ is increased.  The quantum fluctuations of the phase 
of the superconducting wave function are thus described in terms of vortices, 
and in the insulating state (large quantum fluctuations) corresponds to the 
free-vortex state of the $XY$ model~\cite{Fal96}. The mapping 
to the isotropic $XY$ model~\cite{Bra84} can be done in the limit 
$C_0 \gg C$ which results in $K_0=(E_J/8E_{C_0})^{1/2}$, where $E_{C_0}=e^2/2C_0$ 
is the charging energy associated with the stray capacitance of each electrode.  
Thus in this mapping, the junction capacitance, $C$ is neglected.

\begin{figure}
\begin{center}
\epsfig{file=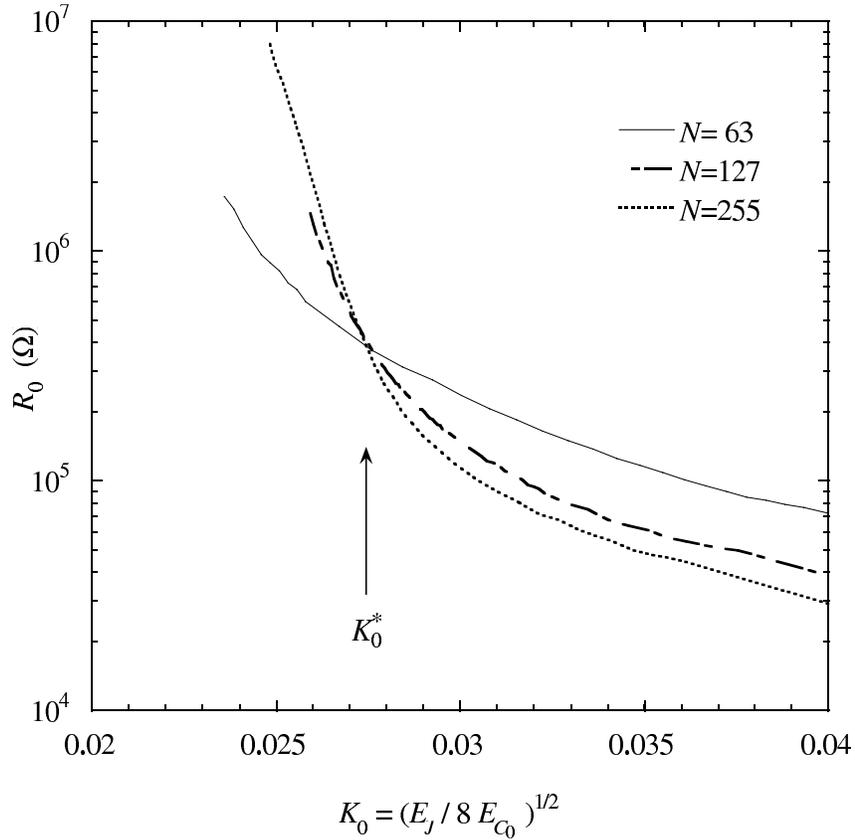,width=0.7\columnwidth,
bbllx=32,
bblly=157,
bburx=564,
bbury=686,
clip=,
angle=0
}
\end{center}
\vspace{-.7\baselineskip}
\caption{Zero-bias resistance taken at $T=0.05$~K 
as a function of the dimensionless coupling constant 
$K_0=(E_J/8E_{C_0})^{1/2}$ for three arrays having different 
number of junctions.
}
\vspace{.3\baselineskip}
\label{fig:Cfig4}
\end{figure}

Figure~\ref{fig:Cfig4} shows a plot of $R_0$ measured at the 
lowest temperature, $T=0.05$~K, vs. $K_0=(E_J/8E_{C_0})^{1/2}$ 
for three arrays with $N=255$, 127 and 63.  
We see in Fig.~\ref{fig:Cfig4} that the three curves for different 
lengths cross at nearly the same point, $K_0^*=0.027$.  To the left of this 
crossing point we have the insulating state, where the resistance is larger 
for the longer arrays.  To the right of this crossing point, we have the 
superconducting state, where the resistance is {\itshape smaller} for longer arrays.  
The arrays in the experiments have $C \gg C_0$, and thus we can not directly 
apply the theory.  In our earlier work~\cite{Cho98} we postulated 
that the effect of $C \gg C_0$ could be accounted for by ``course graining" 
to the scale $\Lambda =(C/C_0)^{1/2}$ , which would result in a coupling constant 
$J=(E_J/\Lambda E_C)^{1/2}$.  Figure~4 of Ref.~\cite{Cho98} shows 
that for this choice of the dimensionless coupling constant, the $R_0$ vs. $J$ 
curves do not cross at the same point, but in the region $J\in\{0.49,0.55\}$.  
Choi {\itshape et al.}~\cite{Choi98} have made a theoretical analysis of the role 
of a finite junction capacitance $C$ by treating $\Lambda$ as a small parameter.  
They found that the transition point should approach the limiting value 
$K_0^c=2/\pi=0.64$ when extrapolated to large $\Lambda$.  The fact that the 
curves cross at one point in Fig.~\ref{fig:Cfig4} would suggest that $K_0
=(E_J/8E_{C_0})^{1/2}$ is indeed the correct parameter for the transition.  
However, the experiment does not support the conclusions of Choi {\itshape et al.} 
in that the experimental critical point $K_0^*\approx0.03\ll K_0^c=2/\pi$.

As we have seen earlier in Fig.~\ref{fig:Cfig3}, 
the resistance at $T>0.6$~K is almost proportional to the array length, 
and in this sense, the arrays behave like classical 1D resistors.  
At lower temperatures, however, large deviations from this classical behavior occur.   
We can qualitatively understand these observations 
in the context of the $(1+1)$D $XY$ model by considering the finite-size 
effect, which plays an important role in a real experiment.  
The zero-bias resistance of the array is determined by quantum fluctuations of 
the superconducting phase, which are described by the vortices in the (1+1)D $XY$
model.  As the temperature is lowered, 
the system becomes larger in the imaginary-time dimension, and at low enough 
temperatures, the system size in the real-space dimension, or the array length, 
determines the energy for free-vortex formation.  The energy increases 
with increasing system size, and thus the probability of free-vortex formation 
is reduced in a longer array.  This means that in a longer array, the superconducting 
state is favored.  If we assume that the zero-bias resistance of the array 
is proportional to the probability of free-vortex formation in the isotropic $XY$ 
model with the area of $N^2$ (i.e., $N$ units in real space and $N$ units 
in imaginary time), we obtain 
\begin{equation}
R_0\sim N^{2-\pi K_0}
\end{equation}
by neglecting any renormalization effects~\cite{GirPC}.  
For $K_0 > K_0^c = 2/\pi$, $R_0$ increases for decreasing $N$, as observed 
in Fig.~\ref{fig:Cfig3} on the superconducting side of the transition.

\section{COULOMB BLOCKADE IN SINGLE JUNCTIONS}
\label{sec:SJ}
\subsection{Theory for current-biased single Josephson junctions}
\begin{figure}
\begin{center}
\epsfig{file=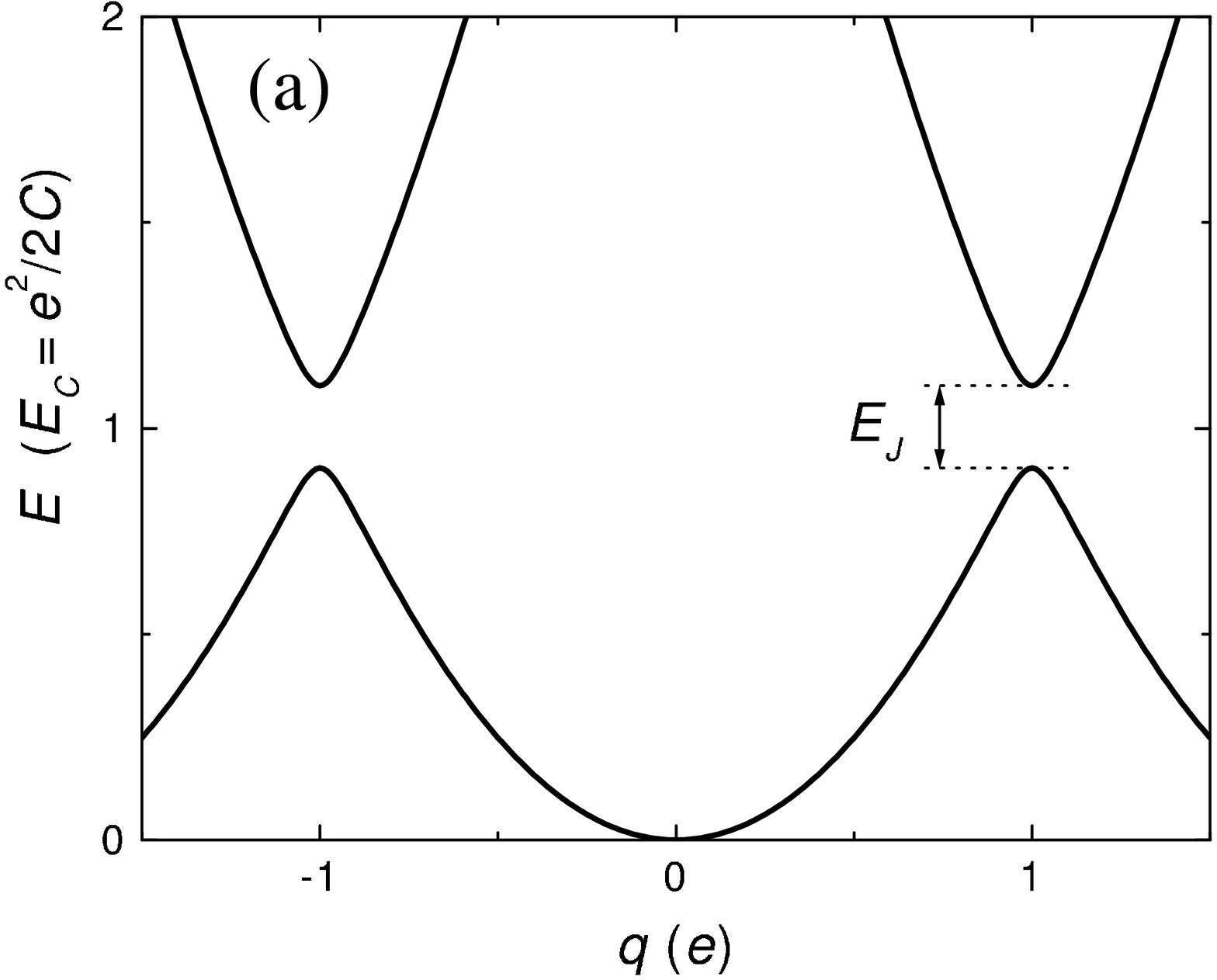,width=.6\columnwidth,
bbllx=18,
bblly=345,
bburx=571,
bbury=793,
clip=,
angle=0
}

\miss\miss

\epsfig{file=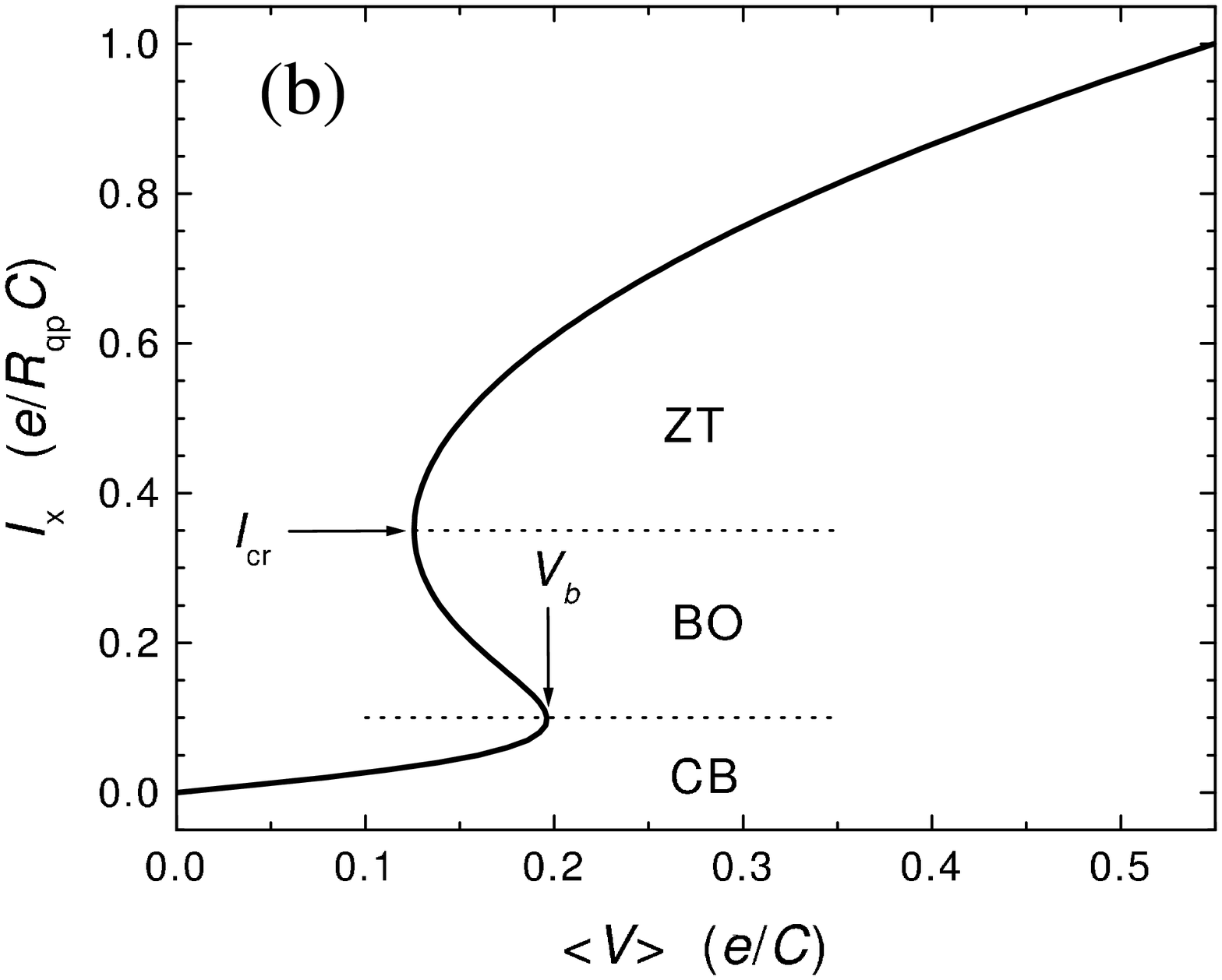,width=.6\columnwidth,
bbllx=17,
bblly=342,
bburx=576,
bbury=791,
clip=,
angle=0
}
\end{center}
\caption
{
(a) Energy diagram and (b) theoretical current-voltage characteristics 
for a single Josephson junction with $E_J/E_C=0.2$ and $R_{\rm qp}
=200\,(h/\pi^2e^2)$ at $k_BT/E_C=0.2$, where $E_J$ is the Josephson 
energy, $E_C\equiv e^2/2C$ is the charging energy, $R_{\rm qp}$ is the 
quasiparticle resistance, and $k_BT$ is the thermal energy.    
(The energy diagram depends only on $E_J/E_C$.)  
}
\label{fig:calc}
\end{figure}

The Hamiltonian of a single Josephson junction in an environment 
with sufficiently high impedance is written as 
\begin{equation}
\label{eq:band1}
H = \frac{Q^2}{2C} - E_J\cos\phi, 
\end{equation}
where $Q$ is the charge on the junction electrode, 
$C$ is the capacitance of the junction, 
$E_J$ is the Josephson energy, and $\phi$ is 
the Josephson-phase difference across the junction.  
The charge $Q$ and $\hbar\phi/2e$ are quantum mechanically 
conjugate valuables, and a set of the eigenfunctions are 
Bloch waves of the form 
\begin{equation}
\psi(\phi)=u(\phi)\exp(i\phi q/2e),    
\end{equation}
where $q$ is called quasicharge 
and $u(\phi)$ is a periodic function, 
\begin{equation}
u(\phi+2\pi)=u(\phi).
\end{equation}
The energy eigenvalue $E$ plotted 
as a function of $q$ has a band structure, 
and in all the allowed bands, it is $2e$ periodic. 
An example of the energy diagram for $E_J/E_C=0.2$, 
where $E_C\equiv e^2/2C$ is the charging energy, 
is shown in Fig.~\ref{fig:calc}a.    
Under constant current bias $I_{\rm x}$, 
in the absence of quasiparticle or Cooper-pair tunneling, 
$q$ increases uniformly in time according to 
\begin{equation}
\label{eq:IV1}
\frac{dq}{dt}=I_{\rm x}
\end{equation}
so that the state of the system advances toward higher $q$ 
within a given band as time goes on.  
The average voltage is given by 
\begin{equation}
\label{eq:IV6}
\langle V\rangle=\sum_{i_b,q}P(i_b,q)\,\frac{\,dE(i_b,q)\,}{dq}\,,
\end{equation}
where $i_b$ is the band index and $P(i_b,q)$ is the probability 
that the system is in the state $(i_b,q)$.  
The probability $P(i_b,q)$ can be calculated by solving  
a set of coupled differential equations of the form 
\begin{equation}
\label{eq:IV8}
\frac{\,dP(i_b,q)\,}{dt}
=\sum_{i_b',q'}A(i_b,q,i_b',q')\,P(i_b',q')=0\,,
\end{equation}
where the matrix element $A(i_b,q,i_b',q')$ describes 
the rate of transition between the states $(i_b,q)$ 
and $(i_b',q')$.  The dominant process for 
$A(i_b,q,i_b',q')$ depends on the magnitude of $I_{\rm x}$.  

An example of the theoretical $I$-$V$ curve is shown in 
Fig.~\ref{fig:calc}b.  
For sufficiently small $I_{\rm x}$ (region~CB), the dominant process 
is stochastic quasiparticle tunneling, where $q$ changes by $e$.  
This tunneling always occurs along the energy parabola $q^2/2C$, 
such that $i_b$ changes by 0 or $+1$ if the initial state is 
in the lowest band ($i_b=1$) and by $\pm1$ 
for all the other initial states ($i_b\geq2$)~\cite{Sch90}.     
The rate for the quasiparticle tunneling
is given by 
\begin{equation}
\label{eq:IV2}
\Gamma(\Delta E)=\frac{\Delta E/e^2R_{\rm qp}}
{\,\exp(\Delta E/k_BT)-1\,}\,, 
\end{equation}
where $R_{\rm qp}$ is the quasiparticle resistance and 
$\Delta E$ is the difference in energy between the initial 
$(i_b,q)$ and final $(i_b',q')$ states, 
\begin{equation}
\label{eq:IV3}
\Delta E\equiv E(i_b',q')- E(i_b,q).  
\end{equation}
At sufficiently low temperatures, the tunneling with 
$\Delta E>0$ is extremely unfavorable, and the $I$-$V$ curve 
is highly resistive (Coulomb blockade).  
For larger $I_{\rm x}$ (region~BO in Fig.~\ref{fig:calc}b), 
the quasicharge is frequently driven 
to the boundary of the Brillouin zone, $q=e$, then taken to $-e$ 
as a Cooper pair tunnels (Bloch oscillation).  
This process decreases $\langle V\rangle$, 
and as a result, the $I$-$V$ curve has a region of negative 
differential resistance, or ``back bending" in the low-current part.  
For still larger $I_{\rm x}$ (region~ZT in Fig.~\ref{fig:calc}b), 
Zener tunneling becomes important, and $\langle V\rangle$ 
increases again.  
In Zener tunneling, no quasicharge is transferred but the state of the 
system jumps from one band to another as it passes by the narrow gap 
between the bands.  The probability of Zener tunneling from band $i_b$ 
to $i_b+1$ or vice versa is given by 
\begin{equation}
\label{eq:IV4}
P_Z=\exp\left[-\frac{\,\pi\,}{8}\frac{\,(\Delta E)^2\,}{\,i_bE_C\,}
\frac{e}{\,\hbar I_{\rm x}\,}\right].
\end{equation}
Following Ref.~\cite{Gei88}
which takes into account the above tunneling processes (quasiparticle, 
Cooper-pair, and Zener), we have calculate the $I$-$V$ curve numerically.     
The parameters for the calculation are $E_J/E_C$, $k_BT/E_C$, 
and $\alpha\equiv h/\pi^2e^2R_{\rm qp}$.  The current and the voltage 
are in units of $e/R_{\rm qp}C$ and $e/C$, respectively.    

As we have seen in Fig.~\ref{fig:calc}b, a typical $I$-$V$ curve 
consists of three regions, so that it is characterized by the local 
voltage maximum, or blockade voltage $V_b$, and the local current 
minimum, or crossover current $I_{\rm cr}$.  
(Here, we have to mention that the back-bending feature is smeared out 
if $k_BT/E_C$ or $\alpha$ is increased considerably.)  
Analytic expression of $V_b$ and $I_{\rm cr}$ has been obtained 
theoretically for limiting cases~\cite{Sch90}.  
The value of $V_b$ is a function of $E_J/E_C$, and given by   
\begin{equation}
V_b \approx 
\left\{
\begin{array}{ll} 
0.25\,e/C\;\; & \mbox{for $E_J/E_C \ll 1$,} \\
\delta_0/e & \mbox{for $E_J/E_C \gg 1$,}
\end{array}
\right.
\end{equation}
as $T\to 0$, where 
\begin{equation}
\delta_0 = \frac{\,e^2\,}{C}\,8\left(\frac{1}{2\pi^2}\right)^{\!1/4}\!
\left(\frac{E_J}{E_C}\right)^{\!3/4}\exp\left[-\left(8\,\frac{E_J}{E_C}
\right)^{\!1/2}\,\right]
\end{equation}
is the half width of the lowest energy band.   
As for $I_{\rm cr}$, 
\begin{equation}
\label{eq:Icr}
I_{\rm cr}\sim\left(I_Z\,\,\frac{e}{\,R_{\rm qp}C\,}\right)^{1/2}
\end{equation} 
is expected for $\alpha\ll(E_J/E_C)^2\ll1$ and $T\rightarrow0$, where 
\begin{equation}
I_Z\equiv\frac{\,\pi\,}{8}\frac{eE_J^{\,\,2}}{\,\hbar E_C\,}
\end{equation}
is the Zener breakdown current.  Note that $I_{\rm cr}$ is much smaller 
than $I_Z$.   
When we compare our experimental results with the theory, we need 
theoretical prediction for finite $k_BT/E_C$, and arbitrary 
$E_J/E_C$ and $\alpha$.  For this reason we have done the numerical 
calculation.  The measured $V_b$ and $I_{\rm cr}$ will be compared 
with the calculation in Sec.~\ref{sec:SJ}.\ref{subsec:comp}.

\subsection{Earier experiments on single junctions}
\label{subsec:EE}
In order to observe the Coulomb blockade in a single junction experimentally, 
the electromagnetic environment for the junction, or the measurement leads 
connected to the junction, should have a high impedance~\cite{Ave91}.  
For this reason, thin-film resistors (NiCr alloy, AuPd alloy, Cr, and 
layered Ge/Pd)~\cite{Hav91} and tunnel-junction (Al/Al$_2$O$_3$/Al) 
arrays~\cite{Gee90,Shi97} were employed for the leads to bias 
a single junction (Al/Al$_2$O$_3$/Al).  

\begin{figure}
\begin{center}
\epsfig{file=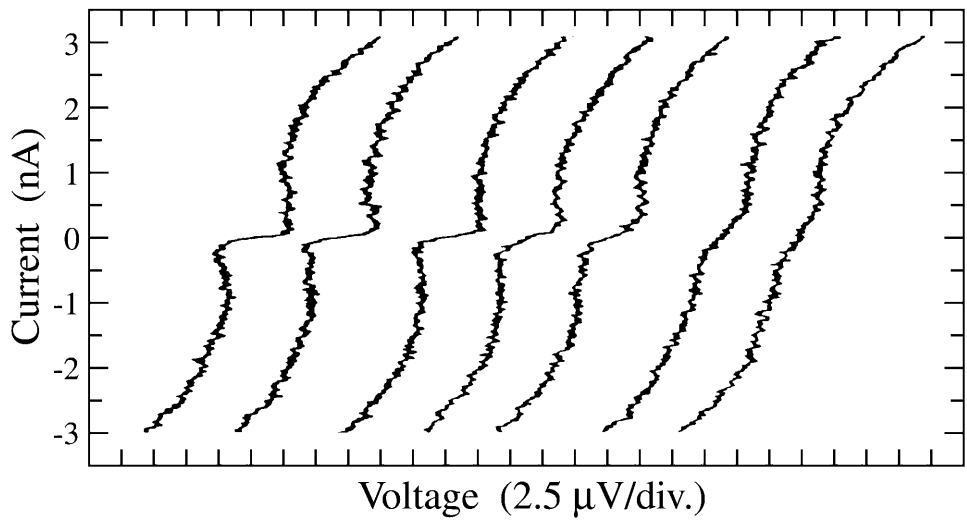,width=0.8\columnwidth
}
\end{center}
\vspace{-.8\baselineskip}
\caption{ 
Current-voltage characteristics of a single Josephson junction 
biased with thin-film resistors at several temperatures.  
From left to right, $T=0.06$, 0.08, 0.11, 
0.17, 0.23, 0.28, and 0.34~K, respectively.  
The origin of the voltage axis is displaced for each curve 
for clarity.  
}
\label{fig:Zfig9}
\end{figure}

Figure~\ref{fig:Zfig9} shows the $I$-$V$ curves of a single Josephson 
junction biased with thin-film resistors at several temperatures.  
A clear Coulomb blockade is seen at $T\leq0.11$~K, and a ``back bending"
is also visible.  In similar samples, high-frequency ($f=0.4-10$~GHz) 
irradiation induced steps in the dc $I$-$V$ curve at $I=\pm2ef$ (not $\pm ef$), 
which can be explained as a phase locking of the externally applied signal 
to the Bloch oscillations~\cite{Hav91}.   

Geerligs used 2D tunnel-junction arrays for the leads~\cite{Gee90}.  
He claimed that in the normal state ($T>T_c$), 
a Coulomb blockade was visible in the single-junction $I$-$V$ curve.  
In the superconducting state ($T<T_c$), however, 
neither the arrays nor the single junction develop clear charging 
effects, and the single junction showed a classic hysteretic 
$I$-$V$ curve with an ordinary supercurrent.  
We believe that his arrays were in the superconducting side of the 
SI transition, and did not have high enough impedance below $T_c$.  

Shimazu {\itshape et al.} biased a single junctions with 1D 
tunnel-junction arrays, and reported an increase of differential resistance 
around $V=0$ in the normal state~\cite{Shi97}.   
In the superconducting state, they measured the zero-bias resistance 
rather than the $I$-$V$ curve.  
The zero-bias resistance in the superconducting state was higher than 
the normal-state resistance for some single junctions, which suggests 
the existence of a Coulomb blockade even in the superconducting state.    
However, this increase of the zero-bias resistance could also be due to 
simple quasiparticle tunneling, which is independent of any Coulomb 
blockade effects for Cooper pairs.   

\subsection{Single Josephson junctions biased with SQUID arrays}
\label{subsec:SJwS}
\begin{figure}
\begin{center}
\epsfig{file=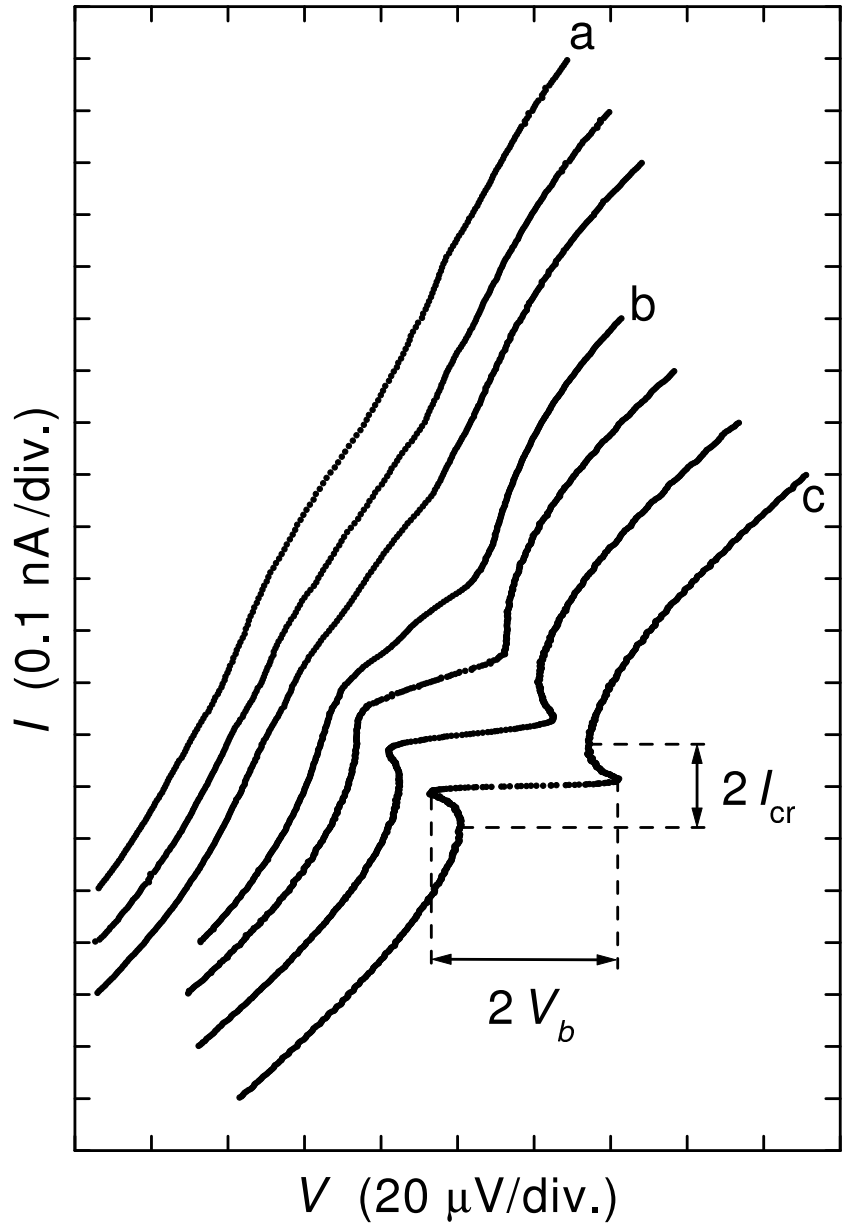,width=0.7\columnwidth,
bbllx=35pt,
bblly=35pt,
bburx=562pt,
bbury=804pt,
clip=, 
angle=0}
\end{center}
\caption{ 
Current-voltage ($I$-$V$) curves of the single junction  
at $T=0.02$~K
for a sample with $r_n=17$~k$\Omega$, $R_n=1.4$~k$\Omega$,  
and $N=65$.  From top left to bottom right, the normalized magnetic field  
$\varphi\equiv BA_{\rm loop}/\Phi_0$ is increased from 0.43 to 0.49 
in steps of 0.01.  The origin of each curve is offset for clarity.  
For the labeled curves, the $I$-$V$ characteristics 
of the leads at the same $\varphi$ are shown in Fig.~\ref{fig:21A4leads}
}
\label{fig:21A4SJ}
\end{figure}

\begin{figure}
\begin{center}
\epsfig{file=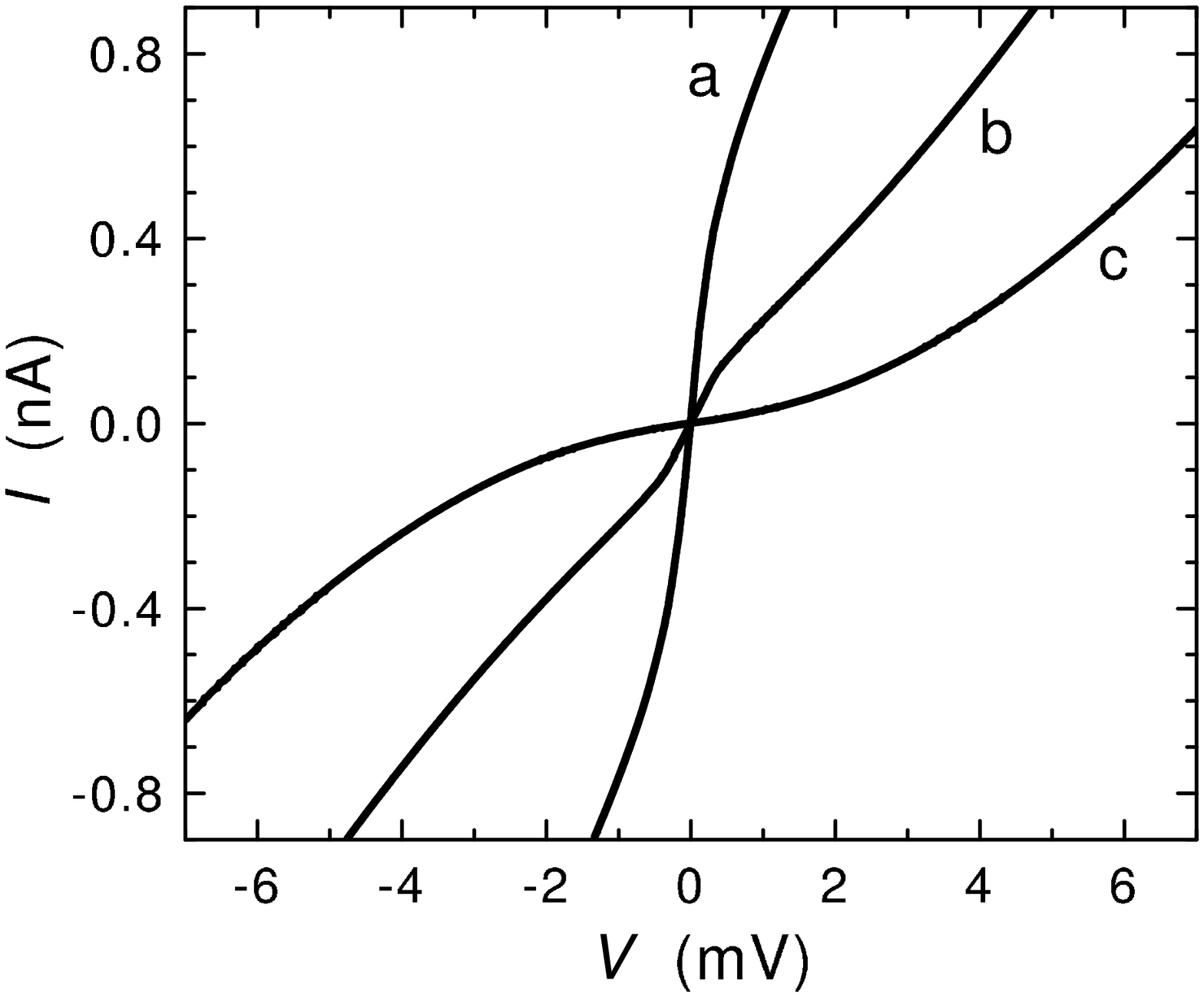,width=0.7\columnwidth,
bbllx=24pt,
bblly=338pt,
bburx=557pt,
bbury=784pt,clip=,
angle=0}
\end{center}
\vspace{-.8\baselineskip}
\caption{ 
Current-voltage curves of the two SQUID-array leads 
connected in series at $T=0.02$~K for the same sample 
as in Fig.~\ref{fig:21A4SJ}.   
From a to c, $\varphi\equiv BA_{\rm loop}/\Phi_0$ 
is 0.43, 0.46, and 0.49, respectively.   
}
\label{fig:21A4leads}
\end{figure}

We have employed 1D arrays of dc SQUIDs for the leads 
to bias a single Josephson junction~\cite{Wat01PRL}.  
The advantage of this SQUID configuration is that in contrast 
to the earlier experiments in Sec.~\ref{sec:SJ}.\ref{subsec:EE}, 
the impedance can be varied {\itshape in situ} by applying an external 
magnetic field at low temperatures.  (See Fig.~\ref{fig:Cfig3}.) 
Thus, we can tune the electromagnetic environment for the single 
junction over a wide range.    
In Figs.~\ref{fig:21A4SJ} and \ref{fig:21A4leads}, 
we show some results on a sample with 
$r_n=17$~k$\Omega$, $R_n=1.4$~k$\Omega$, and $N=65$.  
The $I$-$V$ curves for the single junction at several normalized magnetic fields, 
$\varphi\equiv BA_{\rm loop}/\Phi_0$, are shown in Fig.~\ref{fig:21A4SJ}.  
As $\varphi$ is varied, the $I$-$V$ curve develops a Coulomb blockade.  
We emphasize that the Josephson energy of the single junction 
is independent of $\varphi$, because it does not have a SQUID configuration 
and the field $\varphi\Phi_0/A_{\rm loop}<7$~mT applied here is much 
smaller than the critical field for Al films ($\approx0.1$~T).  
The electromagnetic environment for the single junction 
(the SQUID array), however, is strongly varied with $\varphi$.  
The behavior of the single junction demonstrated in 
Fig.~\ref{fig:21A4SJ} does not result from the 
magnetic-field influencing the single-junction $I$-$V$ curve, 
but rather from an environmental effect on the single junction.  
This experiment demonstrates in a direct way
that the single-junction $I$-$V$ curve is indeed sensitive 
to the electromagnetic environment.  

The $I$-$V$ curves of the two SQUID-array leads connected in series 
at $\varphi=0.43$ ($R_0=0.61$~M$\Omega$), 0.46 ($R_0=3.2$~M$\Omega$), 
and 0.49 ($R_0=43$~M$\Omega$) are shown in Fig.~\ref{fig:21A4leads}.  
The $I$-$V$ curves of the leads are nonlinear, and in general the SQUID 
array cannot be described by a liner impedance model~\cite{Hav00}.  
However, we may characterize the environment by their $R_0$.  
Coulomb blockade is visible only when $R_0\gg R_K$, 
which is consistent with the theoretical 
conditions for the clear observation of Coulomb blockade 
in single junctions~\cite{Ing92}.  
For an arbitrary linear environment characterized by $Z_e(\omega)$, 
$\mbox{Re}[Z_e(\omega)]\gg R_K$ is required for the Coulomb blockade 
of quasiparticle tunneling and $\mbox{Re}[Z_e(\omega)]\gg R_K/4$ 
for that of Cooper-pair tunneling~\cite{Ing92}.  
It is interesting to note that at $\varphi=0.46$ (labeled ``b"), the $I$-$V$ 
curve of the leads is still ``Josephson-like" (differential resistance 
is lower around $V=0$), while that of the single junction is already 
``Coulomb-blockade-like".  This feature becomes more distinct 
in samples with larger $N$~\cite{WatJPCS}.  

\subsection{Comparison with the numerical calculation}
\label{subsec:comp}
\begin{figure}
\begin{center}
\epsfig{file=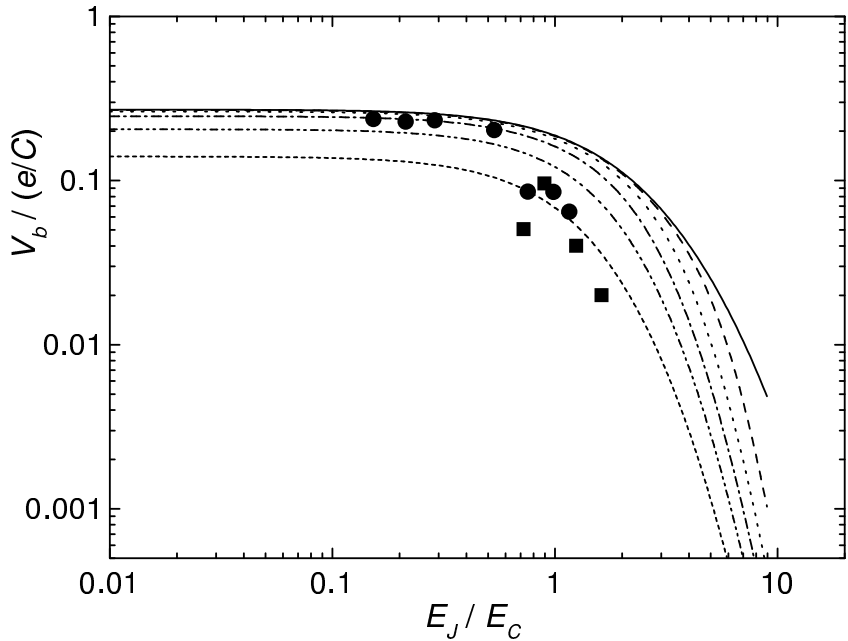,width=0.7\columnwidth,
bbllx=17pt,
bblly=392pt,
bburx=580pt,
bbury=821pt,clip=,
angle=0}
\end{center}
\vspace{-\baselineskip}
\caption{
Blockade voltage $V_b$ divided by $e/C$ as a function 
of $E_J/E_C$. 
From top to bottom, the curves represent 
the numerical calculations for normalized temperatures  
$k_BT/E_C=0$, 0.02, 0.05, 0.1, 0.2, and 0.5, respectively.  
}
\label{fig:Vb}
\end{figure}

The region of negative differential resistance seen 
in Fig.~\ref{fig:21A4SJ} when Coulomb blockade is well developed, 
is related to coherent tunneling of single Cooper pairs 
according to the theory~\cite{Ave91,Sch90} 
of a current-biased single Josephson junction in an environment 
with sufficiently high impedance.  
Following Ref.~\cite{Gei88}, 
we have calculated the blockade voltage $V_b$ 
numerically as a function of $E_J/E_C$~\cite{Wat01SUST}.  
The measured $V_b$ (See Fig.~\ref{fig:21A4SJ}) 
for the samples having nominal 
junction area of $0.1\times0.1$~$\mu$m$^2$ is compared 
with the numerical calculation in Fig.~\ref{fig:Vb}.
The boxes and circles represent the samples biased 
with thin-film resistors (Sec.~\ref{sec:SJ}.\ref{subsec:EE}) 
and with SQUID arrays (Sec.~\ref{sec:SJ}.\ref{subsec:SJwS}), 
respectively.  For the sample shown in Fig.~\ref{fig:21A4SJ}, 
we used the data at $\varphi=0.49$ (curve c) in order to obtain $V_b$.  
At $\varphi=0.49$ the voltage drop at the SQUID arrays is $10^2$ times 
larger than that at the single junction, and the single junction 
is therefore considered to be current biased.  Compare the 
voltage scale of Figs.~\ref{fig:21A4SJ} and \ref{fig:21A4leads}.  
We calculated $E_J$ from $r_n$, $E_J=h\Delta_0/8e^2r_n$.  For $E_C$, 
we employed $c_s=130$~fF/$\mu$m$^2$, and with this value 
the experimental data, especially those for the samples biased 
with SQUID arrays (the circles in Fig.~\ref{fig:Vb}), 
agree with the numerical calculation.  

Actually, a smaller value, $c_s=45\pm5$~fF/$\mu$m$^2$~\cite{Lic89}, 
which was obtained for the junctions with $3\times28$~$\mu$m$^2$ 
and $7\times54$~$\mu$m$^2$, has been frequently 
employed~\cite{Cho98,Hav00,Hav91,Shi97,Hav01}.  
Our apparently large $c_s$ may be partly explained 
by distributed capacitance of the SQUID arrays 
or by residual environmental effects.  
We also note that the uncertainty in $c_s$ 
seems to be large when the junction area 
is on the order of 0.01~$\mu$m$^2$ or smaller.  
For example, 
Fulton and Dolan measured samples with three junctions 
that share a common electrode, and obtained 
$0.20-0.23$~fF for ($0.03\pm0.01$~$\mu$m)$^2\times3$~\cite{Ful87}, 
i.e., $c_s=42-192$~fF/$\mu$m$^2$.  
Geerligs {\itshape et al.} reported $c_s\approx 110$~fF/$\mu$m$^2$ 
for two-dimensional ($190\times60$) junction arrays with 
the areas of 0.01 or 0.04~$\mu$m$^2$~\cite{Gee89}.  
More recently, Penttil\"a {\itshape et al.} studied resistively shunted 
single Josephson junctions with the area of 
$0.15\times0.15$~$\mu$m$^2$~\cite{Pen99}.  
The estimated $C$ of their eight samples ranged between 0.8 and 6.6~fF, 
or $c_s=36-293$~fF/$\mu$m$^2$.  

\begin{figure}
\begin{center}
\epsfig{file=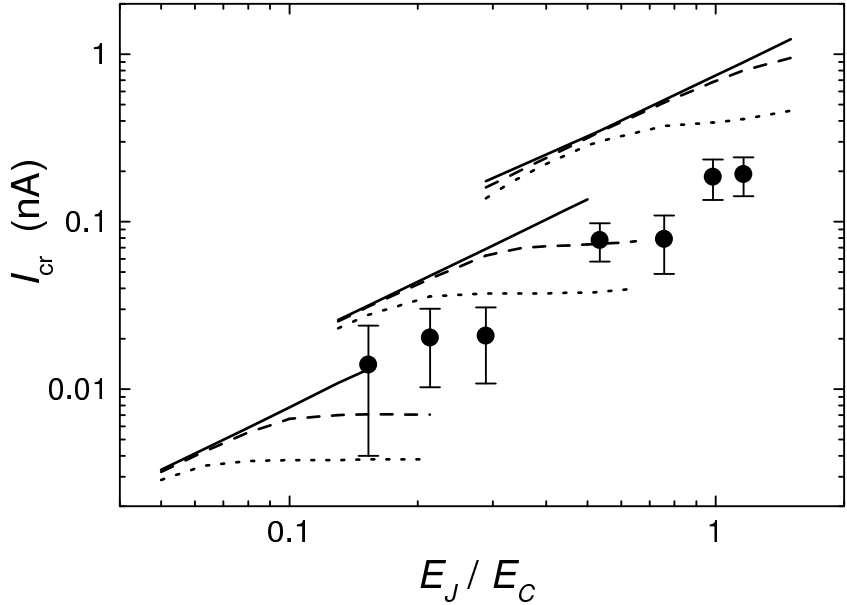,width=0.7\columnwidth,
bbllx=25pt,
bblly=365pt,
bburx=580pt,
bbury=766pt,clip=,
angle=0}
\end{center}
\vspace{-\baselineskip}
\caption{
Crossover current $I_{\rm cr}$ vs. $E_J/E_C$.  
The curves represent 
the numerical calculations for $C=1.3$~fF, and from top to bottom  
$(\alpha\equiv h/\pi^2e^2R_{\rm qp},k_BT/E_C)=(10^{-2},0)$, 
$(10^{-2},0.3)$, $(10^{-2},0.5)$, $(10^{-3},0)$, 
$(10^{-3},0.3)$, $(10^{-3},0.5)$, $(10^{-4},0)$, 
$(10^{-4},0.3)$, and $(10^{-4},0.5)$, respectively.
}
\label{fig:Icr}
\end{figure}

When $C$ is determined, it is possible to estimate $R_{\rm qp}$ 
of the samples from $I_{\rm cr}$.  We plot the measured 
$I_{\rm cr}$ for the single junctions biased with SQUID arrays 
(Sec.~\ref{sec:SJ}.\ref{subsec:SJwS}) as a function of $E_J/E_C$ 
together with some theoretical curves based on our numerical calculation
in Fig.~\ref{fig:Icr}.  
We obtain $R_{\rm qp}=10^0-10^1$~M$\Omega$, or $R_{\rm qp}/r_n 
=10^2-10^3$.  
 
\section{CONCLUSIONS}
One-dimensional (1D) arrays of small-capacitance SQUIDs undergo a sharp 
transition, from Josephson-like behavior to the Coulomb blockade 
of Cooper-pair tunneling, as the effective Josephson coupling 
between nearest neighbors is tuned with an externally applied magnetic field.  
We have shown how length scaling of the zero-bias resistance of the array 
can be used to probe the superconductor-insulator quantum phase transition.  
The observed non-classical dependence of the zero-bias 
resistance on the length of the array, where the zero-bias resistance decreases 
with increasing length, can be supported qualitatively with a theoretical model 
which maps a 1D quantum system to the $(1+1)$D classical $XY$ model.  

We have also used the SQUID arrays as a tunable 
electromagnetic environment for a single small-capacitance 
Josephson junction, and demonstrated how the Coulomb blockade 
of Cooper-pair tunneling is induced in the single junction. 
When the Coulomb blockade is well developed, the measured current-voltage 
curve is consistent with the numerical calculation for a current-biased 
single Josephson junction. 
\section*{ACKNOWLEDGMENTS}
We are grateful to R. L. Kautz for great help in the numerical 
calculation, and to T. Kato and F. W. J. Hekking for fruitful 
discussions.  This work was supported by Swedish NFR, 
and Special Postdoctoral Researchers Program and President's Special 
Research Grant of RIKEN.  The samples for Sec.~\ref{sec:1D} were 
fabricated at the Swedish Nanometer Laboratory.  
M. W. would like to thank the Japan Society for the Promotion 
of Science (JSPS) and the Swedish Institute (SI) 
for financial support.

\end{document}